\documentstyle[preprint,prl,aps]{revtex}
\begin{document}
\draft
\title{Covariant model for dynamical quark confinement}
\author{Helmut Haberzettl}
\address{Center for Nuclear Studies, Department of Physics,\\
The George Washington University, Washington, D.C. 20052}
\date{7 February 1994}
\maketitle

\begin{abstract}
Based on a recent manifestly covariant {\it time-ordered} approach to the
relativistic many-body problem, the quark propagator is defined by a
nonlinear Dyson--Schwinger-type integral equation, with a one-gluon loop.
The resulting energy-dependent quark mass is such that the propagator is
singularity-free for real energies, thus ensuring confinement. The
self-energy integral converges without regularization, due to the chiral
limit of the quark mass itself. Moreover, the integral determines the
low-energy limit of the quark-gluon coupling constant, for which a value of
$g^2/4\pi =4.712$ is found.
\end{abstract}

\pacs{PACS number: 12.38.Aw}
\tighten


In the absence of a direct derivation of the dynamical mechanism of quark
confinement from the QCD Lagrangian, a number of different model approaches
have been considered in recent years to achieve confinement
\cite{stin86,shak89,gogo89,oehm89,tand91,gros91,burd92,buba92}.
(For a recent extensive review of related questions and further
references, see \cite{robe94}.)
Conceptually arguably one of the most elegant mechanisms to
this end is
based on the idea that the quark propagator should not have a pole, a
mechanism which was proposed already twenty years ago \cite{prep74,fuka76}
and which has been implemented in Refs.
\cite{stin86,shak89,gogo89,oehm89,tand91,burd92,buba92}.
In this letter, we want to present a model of quark confinement along
similar lines, i.e., we will derive a quark propagator without any
singularities for real energies.

Our model is based on the cluster-dynamical approach to the relativistic
many-body problem proposed in Refs. \cite{rcd1,rcd2}.
This scattering formulation is a time-ordered one which is manifestly
covariant under an {\it off-shell modification} ${\cal M}$ of the Lorentz
transformation ${\cal L}$. The transformation $(e^{\prime },{\bf q}^{\prime
})={\cal M}(e,{\bf q})$ of an arbitrary four-vector $(e,{\bf q})$ describing
a cluster with three-momentum ${\bf q}$ and off-shell energy $e$ is defined
by relating the three-momenta ${\bf q}^{\prime }$ and ${\bf q}$ in two
different frames by Lorentz transformations involving only the respective
{\it on-shell} energies $\omega ^{\prime }$ and $\omega $, i.e., $(\omega
^{\prime },{\bf q}^{\prime })={\cal L}(\omega ,{\bf q})$, where $\omega
=\omega (m,{\bf q})=(m^2+{\bf q}^2)^{1/2}$, etc., with $m$ being the mass.
The relation between off-shell energies $e^{\prime }$ and $e$ is then
defined by
\begin{equation}
\label{e1}e-\omega (m,{\bf q})=e^{\prime }-\omega (m,{\bf q}^{\prime
})\quad,
\end{equation}
similar to Galilei transformations in Euclidean space. The modified
transformations ${\cal M}$, therefore, leave the differences between on- and
off-shell energies invariant, and reduce to ${\cal L}$ when going on-shell.
Performing then all internal integrations with the appropriate covariant
integral measure [cf. discussion after Eq. (\ref{e6}) below], the
{\it time-ordered} cluster formalism of Refs. \cite{rcd1,rcd2}
is {\it manifestly covariant} under ${\cal M}$ and hence Lorentz-covariant
for arbitrary on-shell matrix elements. We note that, while ${\cal M}$
appears as a nonlinear transformation of four-vectors $(e,{\bf q})$, it can
be can be understood \cite{rcd2} as a covariant four-dimensional
projection of a {\it linear five-dimensional
extension} of the Lorentz transformation ${\cal L}$, i.e., Minkowski space is
a hypersurface of an underlying five-dimensional manifold whose extra,
fifth, dimension is interpreted as the off-shell energy variable with
transformation properties defined in (\ref{e1}), while the usual
Minkowski-space energy component is mapped onto the mass shell.

The simplest possible way of defining a time-ordered quark propagator in the
approach of Refs. \cite{rcd1,rcd2}
is given by the one-gluon-loop expansion depicted in Fig. 1. The
nonlinearity-structure of this integral equation is similar to a
Dyson--Schwinger equation \cite{dyso49,robe94},
albeit in a time-ordered framework. In the center-of-momentum system (CMS),
the resulting dressed (time-ordered) quark propagator is given by
\begin{eqnarray}
t(E-m+i0) & \equiv &
{\displaystyle {{\cal P}(m,{\bf 0}) \over E-m+i0}}\nonumber\\
&   & \nonumber\\
& = &
{\displaystyle {{\cal P}(m,{\bf 0}) \over (E-M+i0)Z-\Sigma (E)}}
\label{e2} \quad,
\end{eqnarray}
where the first identity is a definition of the quark mass $m$; $M$ is the
bare mass, $Z$ the renormalization constant and $\Sigma $ the self-energy
given by the gluon-loop bubble of Fig. 1.
${\cal P}(m,{\bf q})=({\not{\! q}}+m)/2m$
is the covariant {\it on-shell} spin-1/2 projector, i.e.,
$q^\mu=(\omega ,{\bf q})$. (We recall in this context that in
the present formalism \cite{rcd1,rcd2}
{\it all} spin degrees of freedom are described with on-shell energies, as a
consequence of the particular off-shell continuation described above.)
Equation (\ref{e2}) defines the quark mass as a dynamical quantity. As an
invariant, it must be an implicit function of itself, $m=m(E-m)$, where $E-m$
is an invariant according to (\ref{e1}), hence $t(E-m+i0)$ is manifestly
covariant.

In the CMS frame, the explicit energy dependence \cite{note1}
of $m(E-m)$ $\rightarrow m(E)$ is readily found from
(\ref{e2}). Requiring in particular that the mass $m(E)$ be a positive
semidefinite quantity for all energies and using the result (to be shown
below) that $\Sigma (E)\geq 0$ for all energies, it follows that the
renormalization constant $Z$ must be unity, $Z=1$, and hence
\begin{equation}\label{e3}
m(E)-M=\Sigma (E)\quad.
\end{equation}
If we suppose for the moment that $m(E)$ is indeed such that the quark
propagator (\ref{e2}) is free of singularities for real energies, i.e., that
$E-m(E)$ is always negative, then we may introduce a new function $m_0(E)$
by writing $\Sigma (E)=-m_0^2(E)/(E-m)$; we then find
\begin{equation}\label{e4}
m(E)-M=\frac{E-M}2+\sqrt{\left( \frac{E-M}2\right) ^2+m_0^2(E)}\quad.
\end{equation}
For large negative energies, $m(E)$ approaches the bare mass $M$ from above
if we assume that $m_0^2(E)/|E-M|$ goes to zero, i.e., $M$ provides a lower
bound for the dynamical mass $m(E)$. In the following, we will put the bare
mass $M=0$, in other words, we require that the dynamical quark mass
approach its chiral limit for large negative energies,
\begin{equation}
\label{e5}m(E)=\frac E2+\sqrt{\left( \frac E2\right) ^2+m_0^2(E)}\quad.
\end{equation}
Our results will verify that $m(E)$ has indeed the energy behavior given by
this equation.

To this end, we take the quark-gluon vertex to be undressed and given simply
by $g \gamma ^\mu \lambda _a/2$, where $g$ is the coupling constant and
$\lambda _a$ and $\gamma ^\mu $ are the usual color and Dirac matrices,
respectively. From (\ref{e3}), with $M=0$, it then follows, using the rules
of Refs. \cite{rcd1,rcd2}
for evaluating the self-energy bubble $\Sigma (E)$ and keeping in mind the
invariance property (\ref{e1}), that
%
\begin{equation}
m(E) = i\frac 43\frac{g^2}{(2\pi )^4}\int_{-\infty }^{+\infty }
\! de\int \! d^3q\,\frac{m(E-e-\omega +m)}{2q\,\omega (m,{\bf q})}
\,\frac N{(E-e-\omega +i0)\,(e-q+i0)}\label{e6}\quad,
\end{equation}
%
%
where the variables of Fig. 1 were used. Recall here that in the present
approach \cite{rcd1,rcd2}
three-momenta are calculated with on-shell energy parameters in the Lorentz
transformations, and $d^3q\,m/2q\omega $ is the corresponding covariant
integral measure, with normalization factors $2|{\bf q}|\equiv 2q$ and
$\omega /m$ due to the gluon and the quark, respectively. Note that all
masses here are expressed in the CMS notation as explicit functions of the
corresponding off-shell energy; hence, in the integrand the quark mass and
the associated on-shell \cite{note2}
energy are coupled functions of each other and implicit functions of
themselves, i.e., $m=m(E-e-\omega +m)$ and $\omega =\omega {\bf (}
m(E-e-\omega +m),{\bf q}{\bf )}$. The remaining ratio arises from the
energy convolution of the time-ordered quark and gluon propagators whose
detailed variable dependence follows from the invariance requirement (\ref
{e1}). $N$ is the spinor matrix element,
\begin{equation}
\label{e7}N=\bar u({\bf 0},s)\,\gamma _\mu \,\frac{{\not{\! Q}}
+m}{2m}\,N^{\mu \nu }\,\gamma _\nu \,u({\bf 0},s)\quad,
\end{equation}
where $Q^\mu =(\omega ,-{\bf q})$ is the on-shell four-momentum of the
dressed quark in the loop of Fig. 1 and $\bar u$,$\,\,u$ are Dirac spinors
of the quark external to the self-energy loop with third spin component $s$.
$N^{\mu \nu }$ is given by the numerator of the time-ordered gluon
propagator,
\begin{equation}
\label{e8}G^{\mu \nu }(e-q+i0)=\frac{N^{\mu \nu }}{e-q+i0}\quad.
\end{equation}
We treat the gluon here as an undressed particle similar to a bare photon.
The question then arises how to incorporate gluon confinement. This is, of
course, a very complicated and as yet unsolved problem (for recent reviews,
see \cite{robe94,robe93},
and references therein). We deal with it here by what we consider the
simplest possible confinement mechanism known already from QED: discarding
transverse contributions, we restrict the gluons to
longitudinal and scalar (or, temporal)
ones, which are unobservable the same way longitudinal and scalar photons
are unobservable directly. We can offer no deeper justification of this
procedure other than that it works for the present purpose. Hence, in
Feynman gauge one has
\begin{equation}
\label{e9}N^{\mu \nu }=\beta ^\mu \beta ^\nu -\eta ^\mu \eta ^\nu \quad,
\end{equation}
where $\eta ^\mu =(1,{\bf 0})$ and $\beta ^\mu =(0,{\bf n})$, with ${\bf n}$
being the unit vector defined by the gluon three-momentum ${\bf q}=q$\/
${\bf n}$. One finds then that simply $N=-1$.

In evaluating (\ref{e6}), we assume now that there are no singularities of
the integrand in the lower half of the complex $e$ plane other than the one
at $e=q-i0$ (cf. discussion below). The residue of the energy integration
then yields
%
\begin{equation}
m(E) = -\frac{g^2}{3\pi ^2}\int_0^\infty \! dq\,q\,\frac{m(E-q-\omega
+m)}{\omega \left( m(E-q-\omega +m),{\bf q}\right) }\, \frac 1{E-q-\omega
\left( m(E-q-\omega +m),{\bf q}\right) }\label{e10}\quad,
\end{equation}
%
%
where we have also carried out the trivial angle integrations
and dropped the $i0$ in the denominator of the quark propagator in
anticipation of the fact that the mass $m(E)$ will indeed turn out to be
confining. This equation is the main result of the present letter; it
provides a highly nonlinear self-consistency equation for the dynamical quark
mass $m(E)$. Note that, in view of Eq. (\ref{e5}), for large momenta $q$ the
integrand behaves like $m_0^2(-2q)/q^2$, in other words, the integral exists.
[The numerical results discussed below verify that $m_0^2(-2q)$ goes to a
constant for $q\rightarrow \infty $.] The convergence of the integral thus
is achieved {\it by the chiral limit itself}, without any cutoff whatsoever.

Note also that Eq. (\ref{e10}) does not fix any energy scale: using an
arbitrary scale $\Lambda $, the corresponding solution $m_\Lambda (E)$ may
be rewritten as $m_\Lambda (E)=\Lambda \,\mu _1(E/\Lambda )$, where the
functional form of the dimensionless $\mu _1$ is the same as $m_\Lambda $.
Furthermore, $\mu _1$ can be rescaled with an arbitrary (positive,
dimensionless) constant $\lambda $ according to $\mu _1(\epsilon )=\lambda
\,\mu _{1/\lambda }(\epsilon /\lambda )$. This is just the scaling behavior
of the renormalization group. The energy scale must be fixed by other
considerations, e.g., by the value of $m(E)$ at $E=0$, where $m(0)=m_0(0)$,
which should be roughly given by the phenomenological constituent mass
of the quarks,
around 300 MeV, which in turn is about equal to $\Lambda _{QCD}$.

A trivial solution of Eq. (\ref{e10}) is the chiral limit itself, $m(E)=0$.
Since the right-hand side of Eq. (\ref{e10}) is a simple one-dimensional
integral, other solutions can be obtained by an iterative numerical procedure.
In view of the high nonlinearity, however, this is not as easy as it may look
if one would like to find the complete set of solutions \cite{note3}.
At present, therefore, we have made no attempt to
do so but simply solved Eq. (\ref{e10}) by straightforward iteration. As a
starting point, we chose $m(E)$ to be of the form (\ref{e5}) with $m_0$ as a
constant, $m_0=1$, which fixes the energy scale. Both converged solution and
starting function are given in Fig. 2. As can be seen, the solution always
stays above the line given by the pole condition $m=E$. This solution
thus is
indeed confining, approaching the line $m=E$ only asymptotically, as it
must
if quarks are to be free asymptotically. Note also that the starting function
is an excellent approximation of $m(E)$ for negative energies and around
$E=0$. Since Eq. (\ref{e10}) is a nonlinear eigenvalue condition,
the solution of Fig. 2 is found to correspond
to a fixed coupling-constant eigenvalue of
\begin{equation}
\label{e11}\frac{g^2}{4\pi }=4.712\quad,
\end{equation}
a value numerically
accurate to better than 0.1\%. With a straightforward iteration of
Eq. (\ref{e10}), no other eigenvalue could be found. However,
experience with
iterative solutions of other nonlinear equations \cite{note3} does not
rule out the existence of other eigenvalues.

As the low-energy limit of the
QCD coupling constant, the value in (\ref{e11}) is a bit on the low side. We
emphasize, however, that it was obtained without taking into account that the
QCD coupling constant should be a function of the gluon momentum $q$ in (\ref
{e10}). Solving Eq. (\ref{e10}) with a decreasing running coupling constant
$\alpha _s(q^2)$ with a reasonable range of parameters, one finds, first,
that $\alpha _s(0)=g^2/4\pi$ may increase by up to about an order of
magnitude --- which does cover the range of expected low-energy values ---
and, second,
that the corresponding solutions $m(E)$ approach the asymptotic-freedom limit
of $m(E)=E$ faster than for the case shown in Fig. 1. In view of the relative
simplicity of the present model, we do not want to dwell on any details of
this investigation concerning a running coupling constant; nevertheless, we
find it encouraging that the trends go in the right direction.

Our result
verifies that the dynamical quark mass can indeed be written in the form of
Eq. (\ref{e5}). Since everything is completely known in the defining equation
(\ref{e10}), in principle it should be possible to find the analytic
properties of $m(z)$ for an arbitrary complex energy $z$. In practice, this is
complicated very much by the high degree of nonlinearity. Equation (\ref{e5})
suggests that $m(z)$ has square-root cuts in the complex plane with branch
points given by the implicit condition $z=2m(z)$. Various approximations and
numerical studies seem to indicate that this condition cannot be met in the
integrand of Eq. (\ref{e6}) for complex values of $e$ in the lower half of the
complex plane, thus justifying the derivation of Eq. (\ref{e10}). We mention,
however, that we do not consider our findings to be entirely conclusive in
this respect.

In summary, we have presented here the simplest-possible
one-gluon-loop expansion for the quark propagator in the covariant
time-ordered formulation of the relativistic many-body problem of Refs.
\cite{rcd1,rcd2}. We find a confining solution
where the chiral limit of the dynamical quark mass $m(E)$ itself provides the
necessary cutoff to make the self-energy integral convergent.

The author would like to thank Professors D.
Sch\"utte and H. Petry for interesting discussions.
This work was supported in part by the U.S. Department
of Energy, Grant No. DE-FG05-86-ER40270.

\begin{figure}
\caption{Nonlinear one-gluon-loop integral equation
of the time-ordered dressed quark
propagator (solid line with dot) in terms of bare quark propagator
(solid line) and gluon loop (cork-screw line); time proceeds from right
to left. The off-shell energies and three-momenta are given in the
center-of-momentum system of the dressed quark; $e$ and ${\bf q}$ are
the loop-integration variables of Eq. (\protect\ref{e6}).}
\label{fig1}
\end{figure}

\begin{figure}
\caption{Iterative solution of Eq. (\protect\ref{e10}) (solid line).
Both energy
$E\rightarrow \epsilon $ and mass $m\rightarrow \mu $ are mapped here by
the same transformation $x\rightarrow 2\arctan (x/s)/\pi $ from
$(-\infty ,+\infty )$ to the interval $(-1,+1)$; in the graph, the
parameter $s$ is arbitrarily chosen as $s=5$. The energy scale is fixed
by requiring $m(0)=1$ [i.e., $\mu (0)=0.1257$ after transformation].
The long-dashed line is given by Eq. (\protect\ref{e5})
with $m_0(E)$ fixed at
$m_0(E)=1$. The short-dashed straight diagonal line corresponds to the
pole condition $m=E$.}
\label{fig2}
\end{figure}


\begin{references}
\bibitem{note1}For simplicity, we do not use different function names
for the mass depending implicitly on $E-m$ or explicitly on $E$ since
it will be obvious from the context what is meant.
\bibitem{note2}Obviously, if the notion that quarks are confined because
their propagator does not have a pole is to make sense, the word
``on-shell'' has to be taken with a grain of salt. Thus the ``on-shell''
energy for a quark depends on its off-shell energy through the off-shell
dependence of its dynamical mass.
\bibitem{note3}As a rather trivial but still quite illustrative example,
we recall that a quadratic equation $ax^2+bx+c=0$ can be written as
$x=-c/(b+ax)$ and solved by iteration. While, in general, the iteration
will converge very rapidly, it will provide only one of the two
solutions of the quadratic equation, namely the one smaller in
magnitude. To get the larger one, one must iterate
$1/x \equiv y=-a/(b+cy)$.
\end{references}
\end{document}